\documentclass[a4paper,fleqn,usenatbib]{mnras}

\usepackage{newtxtext,newtxmath}

\usepackage[T1]{fontenc}
\usepackage{ae,aecompl}

\usepackage{graphicx}   
\usepackage{amsmath}    
\usepackage{amssymb}    
\usepackage{color}
\usepackage{caption}
\usepackage{subcaption}
\newcommand {\gtau} {g^{(2)}(\tau)}

\title[Temporal intensity interferometry]{Temporal intensity interferometry: photon bunching on three bright stars}

\author[W. Guerin et al.]{
W. Guerin,$^{1}$\thanks{E-mail: william.guerin@inphyni.cnrs.fr}
A. Dussaux,$^{1}$
M. Fouch\'e,$^{1}$
G. Labeyrie,$^{1}$
J.-P. Rivet,$^{2}$
D. Vernet,$^{2}$
\newauthor
F. Vakili,$^{2}$\thanks{E-mail: vakili@oca.eu}
and R. Kaiser$^{1}$
\\
$^{1}$Universit\'e C\^ote d'Azur, CNRS, INPHYNI, France\\
$^{2}$Universit\'e C\^ote d'Azur, OCA, CNRS, Lagrange, France
}

\date{Accepted XXX. Received YYY; in original form ZZZ}

\pubyear{2017}

\begin{document}
\label{firstpage}
\pagerange{\pageref{firstpage}--\pageref{lastpage}}
\maketitle

\begin{abstract}  %
We report the first intensity correlation measured with star light since Hanbury Brown and Twiss' historical experiments. The photon bunching $g^{(2)}(\tau, r=0)$, obtained in the photon counting regime, was measured for 3 bright stars, $\alpha$ Boo, $\alpha$ CMi, and $\beta$ Gem. The light was collected at the focal plane of a 1~m optical telescope, was transported by a multi-mode optical fiber, split into two avalanche photodiodes and digitally correlated in real-time. For total exposure times of a few hours, we obtained contrast values around $2\times10^{-3}$, in agreement with the expectation for chaotic sources, given the optical and electronic bandwidths of our setup. Comparing our results with the measurement of Hanbury Brown \emph{et al}. on $\alpha$ CMi, we argue for the timely opportunity to extend our experiments to measuring the spatial correlation function over existing and/or foreseen arrays of optical telescopes diluted over several kilometers. This would enable $\mu$as long-baseline interferometry in the optical, especially in the visible wavelengths with a limiting magnitude of 10.
\end{abstract}




\section{Introduction}

Optical aperture synthesis interferometry on baselines as long as $300$~m,
\emph{i.e.} $1$~mas resolution, has become a routine technique for
high angular resolution imaging, mostly of stellar objects with the
VLTI, the CHARA Array, and the NPOI~\citep{Mourard:2015,Garcia:2016,Gomes:2017}.
In the next decade extremely large telescopes with $30-40$~m apertures using full-fledged multi-conjugate adaptive optics and laser-guide stars, together with multi-aperture amplitude interferometers, will cover the angular resolution range from $1$~as (seeing limited) to $1$~mas over a spectral range from $0.6$ to $10$~$\mu$m. The domain of even higher resolution, down to $10$~$\mu$as, remains a real challenge on the long term as a large number of telescopes separated over kilometers must be co-phased by fringe tracking at a millisecond timescale in the visible. The situation gets even more stringent for faint and/or highly resolved objects that constitute in fact the most interesting scientific targets.

Intensity interferometry is an alternative solution to reach high angular resolution. As demonstrated by Hanbury Brown and Twiss~\citep{HBT:1956}, HBT hereafter, it is based on measuring the intensity correlation of light detected by two (or more) spatially separated telescopes pointing the same star, with zero time delay, $g^{(2)}(\tau=0, r)$. The angular diameter of the star is the first quantity which can be extracted from this correlation function as a function of increasing baselines $r$, before accessing to more details on its surface structures. The main advantage of this type of interferometer is that it only requires to control the light path over a fraction of the detection time resolution (a few cm for a time resolution of 500\,ps), implying a robustness against atmospheric turbulence, even in the visible, and allowing its application to long baselines and thus high angular resolution. The astronomical observations made with the HBT interferometer in the 1960's and 1970's in Narrabri allowed to extract the angular diameter of more than 30 stars with a resolution better than 1~mas ~\citep{HBT:1974}. But, after the 1970's Narrabri astronomical observations, intensity interferometry has been abandoned in astrophysics (it is still widely used in particle physics~\citep{Alexander:2003}), mainly due to its inherent shortcomings in terms of sensitivity, which implied, at that time, large-area collectors (several meters in diameter) and very long integration times (several hours or even tens of hours)~\citep{HBT:1968}.

However, during the last decades, the continuous and very significant progress in photonic technologies and related digital electronics has opened new perspectives for HBT interferometry. This technique could be considered nowadays as a realistic solution for very long baseline optical interferometry, for example in the $10$~km range, based on the use of Cherenkov arrays of photon buckets~\citep{LeBohec:2006,Dravins:2008b}. The growing interest to revive intensity interferometry has thus triggered a number of simulations and laboratory experiments~\citep{Naletto:2009,Nunez:2012,Rou:2013,Dravins:2014,Dravins:2015a,Dravins:2015b,Dravins:2016,Sinclair:2016,Pilyavsky:2017}, considering even the third-order intensity correlation with three separate telescopes, to retrieve the phase of the Fourier transform of the science object under study~\citep{Nunez:2015}. In this context, photon bunching of Sun light has been recently observed in~\citep{Tan:2014,Tan:2016}, the light being collected by one aspheric lens, thus equivalent to one telescope. Further attempts have been made to detect HBT signal on stars other than the sun, either on the VERITAS array~\citep{Dravins:2014} or from rather modest 1-2~m class telescopes, separated by 50~m at Lowell~\citep{Horch:2013} and 3.9~km in Asiago~\citep{Zampieri:2016}. But, to our knowledge, no significant $g^{(2)}$ signal from stars other than the sun has been reported since the pioneering work by HBT. The main reason is the poor signal to noise ratio (SNR). Whereas Tan and coworkers~\citep{Tan:2014,Tan:2016} have demonstrated the possibility to obtain a SNR greater than one on the sun, taking advantage of the high photon flux, which allows the use of monomode fibers and narrow filters, this is far more difficult for other stars.

The main technical requirement to perform intensity-correlation measurements is the ability to detect $g^{(2)} (\tau=0, r=0) > 1$ (``photon bunching'') with a good SNR. This can be tested with a single telescope by measuring the temporal autocorrelation function for the intensity $I$,
\begin{equation}\label{eq.g2}
g^{(2)} (\tau, r=0) = \frac{\left\langle I(t) I(t+\tau)\right\rangle}{\left\langle I(t) \right\rangle^2} \, ,
\end{equation}
noted $\gtau$ hereafter. Furthermore, beyond the simple test of the SNR, this configuration also gives some important informations about the light emitted by stars. In particular, the value of $g^{(2)} (\tau=0)$ depends on the emission process, allowing the classification of different types of sources. For example, for chaotic light one has $g^{(2)} (\tau=0) = 2$, whereas for laser emission one gets $g^{(2)} (\tau=0) = 1$~\citep{Loudon:book}. While intensity interferometry in the spatial domain is promising to increase the angular resolution, it also opens, in the temporal domain, the route to the observation of emission processes such as non classical phenomena in astrophysical sources ~\citep{Johansson:2007,Dravins:2008a,Foellmi:2009,Tan:2014}.

In this paper, we report temporal photon bunching on stars other than the sun. To our knowledge, this constitutes the first successful on-sky intensity-correlation experiment since HBT, and first ever in the photon-counting regime. Moreover, this is performed with a telescope of only 1~m in diameter. The paper is organized as follows. Section \ref{sec.setup} presents the experimental setup for measuring the intensity correlation function. In Sec.\,\ref{sec.LabTest} we compute the expected contrast and we report on the tests and calibrations performed in the laboratory using an artificial star based on a thermal light source. Section \ref{sec.results} presents the on-sky experiments and the results obtained on three real stars. Finally we summarize our results and discuss possible future developments.

\section{Experimental setup}\label{sec.setup}

The experimental setup for measuring the intensity correlation function $\gtau$ is composed of an optical coupling device, to filter and transport light, and a photon-counting device, for detection, digitization, and signal processing.

\subsection{Optical coupling device}\label{sec.optics}

\begin{figure*}
    \begin{subfigure}{\columnwidth} 
        \centering \includegraphics[width=\columnwidth]{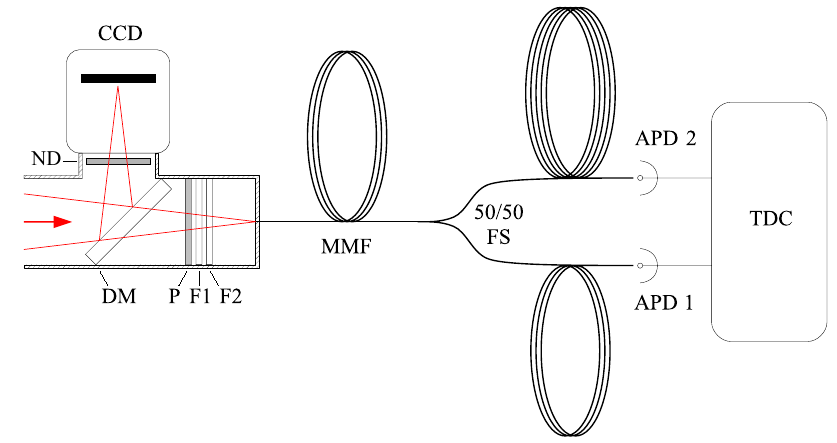}
        \caption{Scheme of the optical setup to filter the
light, inject it into a multimode fiber (MMF) and detect it. The
light is coming from the left, either from the telescope
(on-sky measurements) or from a point source imaged with a
lens on the fiber tip (laboratory test). DM\,: Dichroic beam splitter.
P\,: Polarizer. F1\,: $100$~{\AA} bandwidth filter. F2\,: $10$~~{\AA} bandwidth
filter (both filters centered on $7800$~{\AA}).
CCD\,: Guiding CCD camera to monitor the star position.
ND\,: neutral density (1/10). FS\,: 50:50 Fiber
splitter. APD\,: Avalanche photodiodes. TDC\,: Time to digital converter.}\label{fig.setupa}
    \end{subfigure}
    \hspace{0.5cm}
    \begin{subfigure}{\columnwidth}
        \centering \includegraphics[width=\columnwidth]{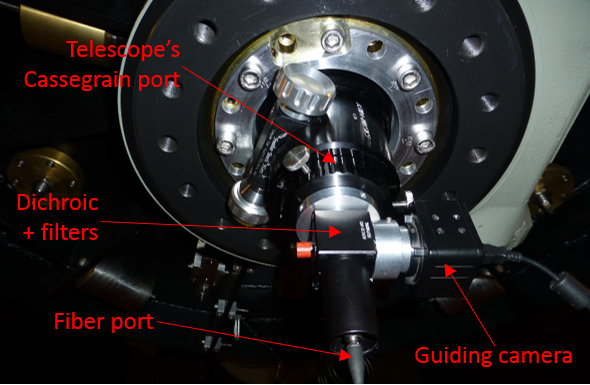}
        \caption{Coupling device connected to
the telescope's Cassegrain port through a standard 2-inch eyepiece holder, in which two cascaded focal reducers are placed (not visible).}\label{fig.setupb}
    \end{subfigure}
    \caption{Experimental setup}\label{fig.setup}
\end{figure*}

In order to minimize mechanical constraints and for ease of use on the telescope, we have chosen to employ fiber technology to transport the light from the source (artificial or telescope focus) to the detectors. Before detection, the light has to be spectrally filtered in order to increase the coherence time and thus enhance the contrast of the correlation (see Section~\ref{sec.calculs}). We have chosen to design a compact optical setup, as depicted in Fig.~\ref{fig.setupa}, without any moving parts, which can be easily transported and used for laboratory tests as well as set up at the telescope focus.

The first optical element is a dichroic plate that reflects light for $\lambda < 6500$~{\AA} and transmits it for $\lambda > 6500$~{\AA}. The reflected light is used to monitor the star's position with a guiding CCD camera. The transmitted light is used for the intensity correlation measurement. The transmission of the dichroic plate at our working wavelength $\lambda_0 = 7800$~{\AA} is 97.5\%. The light then goes through a series of $1$-inch optical elements before reaching the fiber input. First, a linear polarizer is used to restrict the state of the light to one polarization mode, which filters out half of the signal. Then, the light is spectrally filtered by a narrow-band dielectric filter with a full width at half maximum (FWHM) bandwidth of $\Delta \lambda = 10$~{\AA} centered at $\lambda_0$. Since this filter does not completely filter out all the light in the UV and IR, we add a second dielectric band pass filter centered at $\lambda_0$ with a larger bandwidth of 100~{\AA}. From the specifications of the filters, we estimate the relative off-band transmission to be below 0.2\%, negligible for our measurements. The total transmission at $\lambda_0$ of the combined filters has been measured to 61\%, in agreement with the specifications. The total peak transmission at $\lambda_0$ of the complete setup is thus estimated to $\sim 30\%$.

The fiber used to transport light from the source to the detection device is a 20~m-long multimode graded-index fiber (MMF) with a core diameter of 100~$\mu$m and a numerical aperture of 0.29. It is connected to a 50:50 fibered splitter such that two photodetectors can be used to compute the autocorrelation function without being limited by the dead time $\tau_\mathrm{d} \simeq 22$~ns of each detector, much larger than the coherence time, $\tau_\mathrm{c} \sim \lambda_0^2/(c \Delta\lambda) \simeq 2$~ps. The two output ports of the splitter are connected to the photodetectors via two other MMFs of $1$~m for the first detection channel and $2$~m for the second.

\subsection{Photon-counting device}\label{sec.electronics}

The photons are detected by two fiber-coupled avalanche-photodiodes (APDs) in the photon-counting regime. Their typical detection efficiency at $\lambda_0 = 7800$~{\AA} is $64\%$ on an active area  of diameter 180~$\mu$m. An important parameter is their temporal resolution, which will limit the temporal width of the measured photon bunching peak. The manufacturer's specification gives a jitter function of FWHM $350$~ps for a 10~$\mu$m diameter light spot, but it should be degraded when using 100~$\mu$m MMFs, such that 500~ps is a realistic expectation. Their dark count is on the order of $100$ counts per second (cps), well below the count rate in our measurements, which is on the order of $10^6$~cps. The APDs are placed in thermo-regulated boxes. In order to avoid any spurious correlation induced by electronic cross-talk, we also introduce an electronic delay between the two channels by using a 10~m shielded BNC cable for the second channel and 1~m for the first channel. The total delay (optical and electronic) between the two channels is $t_0 \simeq 45$~ns and is subtracted in the presented data.

For each detected photon, the APDs produce a 10~ns pulse, whose rising edge is detected and processed by a time-to-digital convertor (TDC) with a time bin of 162~ps. The TDC is operated using the ``Delay-Histogram'' mode, which allows computing the histogram of the number of coincidence $N$ per time bin between the two detectors, and thus the intensity correlation function
\begin{equation}\label{eq.g2}
g^{(2)} (\tau) = \frac{\left\langle N(t) N(t+\tau)\right\rangle}{\left\langle N(t) \right\rangle^2} \, ,
\end{equation}
with a fast and easy data processing. The data transfer rate of the TDC is however limited to $5\times 10^6$~cps by the USB link to the computer.

\subsection{Data acquisition} \label{sec.TDC}

Our TDC (model ID801 by ID Quantique) can be operated in two different acquisition modes. In the ``time-tagging'' mode, the arrival time of each photon is recorded with the corresponding channel number, and the data can be processed afterwards to compute the $\gtau$ function. In that case, a full night of acquisition would produce a huge amount of data and would require a lengthy data processing~\citep{Zampieri:2016}. Processing the data in real time is possible at the cost of reducing the duty cycle of the measurement, dividing the useful observation time by $\sim 3$ (depending on the photon flux). Moreover, in this mode the data transfer rate of the TDC is limited to $\sim 2.5\times 10^6$~cps for all channels, which is not appropriate for the bright stars we observed. We thus used the ``Delay-Histogram'' mode, which allows computing the histogram of the delays between the arrival time of the photons on the two channels in real time, in a range between 0 and 64~ns. With this mode the data transfer of the TDC is limited to $\sim 5\times 10^6$~cps, and the data processing and saving is straightforward.

However, the delay histogram is not strictly equivalent to the $\gtau$ function, which is the normalized histogram of the delays between \emph{all} pairs of photons arriving on channel one and two. On the contrary, for each photon arriving on channel one (a START photon), the delay histogram takes into account only the first next photon arriving on channel two (STOP). The probability of detecting long delays is thus lower, and the histogram has to be normalized by an exponential decaying function. Moreover, if a second START photon is detected before the STOP, only the shortest delay is registered, which means that the first START photon is not used. This acquisition mode is thus not the one that optimizes the SNR of the measurement, since many photons are unused in the data processing.

One limitation of TDCs is the fact that all time bins do not have the same exact duration. The variation is however not random, but has a periodic pattern, which induces a correlation artefact, on the order of 1\%. This spurious, repeatable correlation must be subtracted from the measurement. This is achieved by acquiring a histogram with a completely uncorrelated light, using an incandescent light bulb without any filter (white light) such that the coherence time is too small to induce any measurable correlation. This histogram $H_\mathrm{white}$ is subtracted from the data histogram $H_\mathrm{data}$ after normalization and we finally compute
\begin{equation}
\gtau = 1 + H_\mathrm{data}-H_\mathrm{white} \; .
\end{equation}
The ``white'' histogram is taken in similar conditions as the data and with a very long integration time such that its noise is small compared to the noise of the data.

\section{Theoretical expectation and laboratory test}\label{sec.LabTest}

\subsection{Expected contrast}\label{sec.calculs}

For chaotic light (thermal source), the intensity correlation function is given by~\citep{Loudon:book}
\begin{equation}\label{eq.siegert}
g^{(2)} (\tau) = 1 + \beta |g^{(1)} (\tau)|^2,
\end{equation}
where $g^{(1)} (\tau)$ is the first order (field-field) correlation function, which can be computed by the Fourier transform of the optical spectrum, and $\beta = 1/M$ is a factor accounting for the loss of coherence due to the number $M$ of detected spatial and polarization modes. With polarized light and a point source (unresolved star), $M=1$.
The theoretical value at zero delay is thus $g^{(2)}(0)=2$, i.e. the contrast, defined as $\mathcal{C} = g^{(2)}(0)-1$, is 1 for a spatially coherent thermal source. However, to measure such a contrast, one needs an electronic setup fast enough to resolve the temporal coherence time, inversely proportional to the spectral width. This can be done with pseudo thermal light, for example based on a rotating ground glass disc~\citep{Arecchi:1966} or Doppler broadening of light scattered by laser-cooled atoms, with microsecond timescale~\citep{Nakayama:2010}. High contrast with subnanosecond resolution has also been obtained using light scattered by atomic vapor at room-temperature~\citep{Dussaux:2016}.

To be able to observe photon bunching from a real thermal source, one needs to reduce the extent of the spectrum (i.e. increase the temporal coherence time $\tau_\mathrm{c}$) using filters. In our current system, $\tau_\mathrm{c}$ is fixed by the narrowest filter, which has approximately a rectangular 10\,{\AA}-wide transmission spectrum, corresponding to a frequency width of $\Delta \nu \simeq 500$\,GHz. This gives a temporal coherence function close to $\gtau \simeq 1 + $sinc$^2(\Delta\nu\tau)$ with a FWHM $\tau_\mathrm{c} \simeq 1.6$\,ps. This temporal coherence time is much shorter than the 500\,ps time resolution of the APDs, inducing a reduction of the measured contrast as well as a broadening of the correlation function. The expected intensity correlation function is the convolution between the ideal correlation function and the time response of the photodiodes. This leads to an expected contrast $\mathcal{C} \simeq 2.2\times 10^{-3}$ with a FWHM given by the relative jitter of the two APDs, i.e. $\sqrt{2}\times 500 \simeq 700$~ps.

\subsection{Test with an artificial star}\label{sec.lab}

\begin{figure}
    \includegraphics[width=\columnwidth]{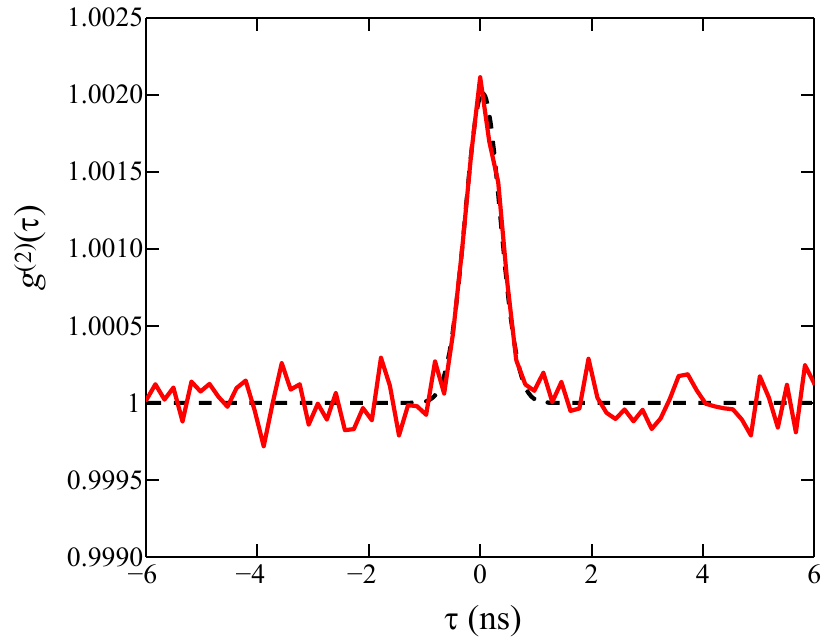}
    \caption{Temporal intensity correlation function $\gtau$ measured on a thermal source in the laboratory with our optical and electronic setup. The integration time is $70.5$~h with an average count rate of $1.42\times 10^6$~cps per detector.}
    \label{fig.bunching_lab}
\end{figure}

In order to test our optical and electronic setup, we have performed laboratory measurements using a light source that acts as an artificial star. To produce a spatially coherent broadband light, we have injected light from a simple light bulb into a single-mode fiber. The output of the fiber is then focused by a lens into our optical filtering system with an aperture that approximately mimics the aperture of the telescope that we used on the sky.

The result of the temporal intensity correlation function $\gtau$ measured with this artificial star is reported in Fig.\,\ref{fig.bunching_lab}. Here we have used of very long integration time (70.5 h) to achieve a very low noise. The bunching peak at $\tau \simeq0$ is well visible. A Gaussian fit to the data yields a measurement of the contrast $\mathcal{C} = (2.01 \pm 0.09) \times 10^{-3}$ ($1\sigma$ confidence interval), and a FWHM width of $740 \pm 35$~ps, in agreement with the evaluations made in Section \ref{sec.calculs}.

This test validates our experimental setup as well as our data acquisition and analysis procedure.

\section{Temporal correlation on stars}\label{sec.results}

\subsection{Telescope}\label{sec.telescope}

We used the $1.04$-meter West Cassegrain telescope
of the C2PU facility (Observatoire de la C{\^o}te d'Azur, OCA, Calern Plateau site,
UAI code 010, latitude\,: $43^\circ\,45'\,13''$ North, longitude\,:
$6^\circ\,55'\,22''$ East, altitude\,: $1270$~m).
This telescope has two possible optical configurations\,: a F/3.2 prime focus with 3-lenses Wynne corrector and a F/12.5 pure Cassegrain focus. For our experiment, we have chosen the Cassegrain configuration (see Fig.~\ref{fig.setupb}).
The central obscuration due to the Cassegrain secondary mirror (including the 4-vane spider) is $9.7\%$ of the entrance pupil surface.
The telescope is on an equatorial yoke mount. Consequently, circumpolar targets are out of scope and the maximum accessible declination is $+60^\circ$.

\begin{table*}
\caption{Main characteristics of the three stars used for this experiment.}
\label{tab.Stars}
\begin{tabular}{lllllll}
\hline
Name & $\alpha_{2000}$ & $\delta_{2000}$ & R mag & I mag & Spectral type & Comment \\
\hline
$\alpha$ Boo$^\star$  & $14^h\,15^m\,39.67^s$ & $+19^\circ\,10'\,56.7''$ & $-1.03$ & $-1.68$ & K0-III C & RGB star \\
$\alpha$ CMi & $07^h\,39^m\,18.12^s$ & $+05^\circ\,13'\,30.0''$ & $-0.05$ & $-0.28$ & F5IV-V & Spectroscopic binary$^\dagger$\\
$\beta$ Gem & $07^h\,45^m\,18.95^s$ & $+28^\circ\,01'\,34.3''$ & $+0.39$ & $-0.11$ & K0-III & High proper motion star \\
\hline
\end{tabular}
\\ \raggedright \footnotesize{$^\star \alpha$ Boo is very marginally resolved with a 1~m telescope: the uniform disk diameter in I band being 20~mas, the visibility factor is above $0.96$. $^\dagger\alpha$ CMi companion is a white dwarf, which does not affect the temporal bunching.}
\end{table*}

\begin{table*}
\caption{Main circumstances for the observing runs performed on the three stars. Begin and end dates are in UTC (ISO~8601 compact format). $a$ is the air mass range and the seeing is given as a median value.}
\label{tab.ObsRuns}
\begin{tabular}{lllll}
\hline
Star& Begin & End & $a$ & Seeing\\ 
\hline
$\alpha$ Boo \, & $20170220T230513Z$\, & $20170221T053214Z$\, & $1.10\rightarrow2.15$\, & $1.43''$\\ 
$\alpha$ Boo \, & $20170221T232206Z$\, & $20170222T051625Z$\, & $1.10\rightarrow1.92$\, & $1.25''$\\ 
$\alpha$ CMi \, & $20170220T205428Z$\, & $20170220T230512Z$\, & $1.28\rightarrow1.45$\, & $1.43''$\\ 
$\alpha$ CMi \, & $20170221T194301Z$\, & $20170221T232205Z$\, & $1.28\rightarrow1.52$\, & $1.25''$\\ 
$\beta$ Gem \, & $20170222T184704Z$\, & $20170223T021135Z$\, & $1.04\rightarrow2.09$\, & $2.14''$\\ 
\hline
\end{tabular}
\end{table*}

The native Cassegrain focus has a focal length of $13$~m. So,
an astronomical observation conducted in median seeing conditions
($1.5''$) would lead to a stellar Point Spread Function (PSF) with $95$~$\mu$m FWHM.
However, to improve the coupling efficiency in average
observing conditions (no real-time correction of the turbulent
tip-tilt and of fast tracking residual defects), we have chosen to reduce the
aperture ratio to F/5.6 by cascading two standard focal reducers.
This reduced aperture ratio remains compatible with the fiber numerical aperture of $0.29$. The resulting field of view on the iNova PLB-Mx guiding CCD camera is then $2.8'\times 2.1'$ and the PSF has a FWHM as small as
$42$~$\mu$m for a seeing of $1.5''$. This is comfortably smaller than the fiber's core diameter of $100$~$\mu$m.

Since only very bright stars were used for this preliminary experiment, we had to introduce a neutral $1/10$ density on the guiding beam (after the reflection on the dichroic plate), to avoid saturation on the guiding sensor (Fig.~\ref{fig.setup}). Both manual and automatic guiding corrections were used. However, sidereal tracking drifts are slow enough to make manual correction sufficient.

\subsection{Observation conditions}\label{sec.observations}

The measurements were carried during the three nights
20, 21, and 22 February 2017 (respectively 6, 5, and 4 days before the
New Moon). For the two first nights, the seeing ($1.43''$ and $1.25''$
respectively) and scintillation ($3.7\%$ and $3.3\%$ respectively), can be considered as average compared with usual values on this observing site.
For the third night however, the median value of the seeing ($2.14''$)
can be considered as poor. The seeing and scintillation measurements
are provided by the G-DIMM instrument~\citep{Aristidi:2014}
of the local turbulence monitoring station.

Three very bright stars were used\,:
$\alpha$ Boo (Arcturus), $\alpha$ CMi (Procyon), and $\beta$ Gem (Pollux).
Table~\ref{tab.Stars} sums up basic data about these sources, and Table~\ref{tab.ObsRuns} gives the circumstances for the observing runs performed on these stars.

\subsection{Results}\label{sec.g2_results}

A first important result is to quantify the photon flux that we are able to inject into the multimode fiber. With the brightest star we observed ($\alpha$~Boo), we could obtain photon-counts larger than $5\times10^6$~cps per detector. Consequently, we had to set the telescope slightly out of focus to avoid the saturation of the data transfer link from the TDC to the computer. With $\alpha$~CMi, the average photon-count over the whole integration time is $\mathcal{F} = 1.44\times10^6$~cps per detector. Given the quantum efficiency $\eta=0.64$ of the detectors, this amounts to a total photon flux of $4.5 \times 10^6$~s$^{-1}$ at the outlets of the two fibers.
The photon rate at the input of the fiber can be estimated as
$6.8\times10^6$~s$^{-1}$. This takes into account {\sl (i)} the magnitudes of
 the star in $R$ and $I$ bands, {\sl (ii)} a modelled atmospheric absorption
at $780$~nm for a standard winter air at $1270$~m elevation, and an effective
air mass of $1.32$, {\sl (iii)} the collecting area of the telescope
($7729$~cm$^2$) and a \emph{bona fide} estimate ($44\%$) of losses on
the mirrors and focal reducers, {\sl (iv)}
the characteristics of the beam splitter, polarizer and filters (see Section~\ref{sec.optics}). This yields
an injection efficiency close to $66\%$, which accounts for the coupling
at the inlet of the fiber, and the losses along the fiber system (fibers and splitter) itself.
This efficiency fluctuates, depending on the seeing conditions. For the observation of $\beta$ Gem, the atmospheric conditions were poor and the efficiency was indeed lower.

\begin{figure*}
    \begin{subfigure}{0.66\columnwidth} 
        \centering \includegraphics[width=\columnwidth]{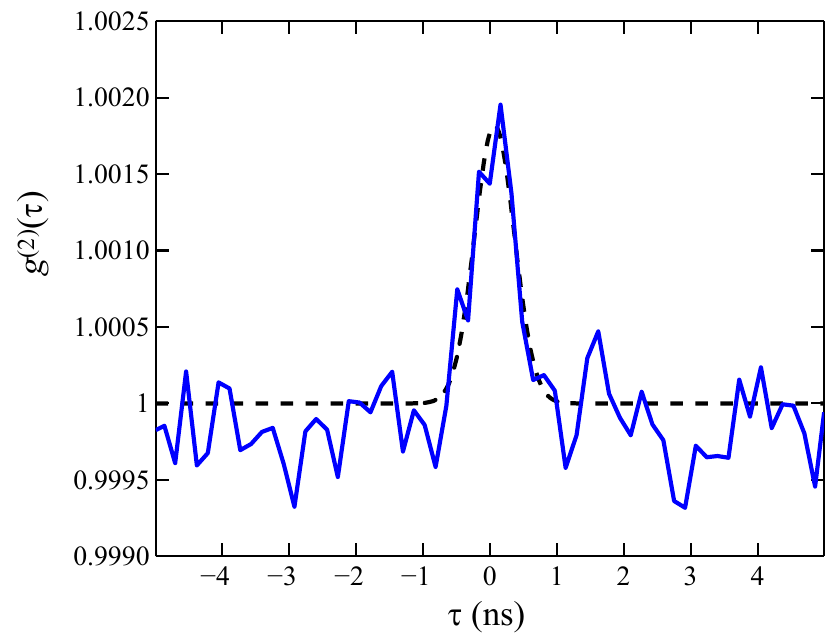}
        \caption{$\alpha$ Boo (Arcturus).}\label{fig.g2a}
    \end{subfigure}
    \begin{subfigure}{0.66\columnwidth}
        \centering \includegraphics[width=\columnwidth]{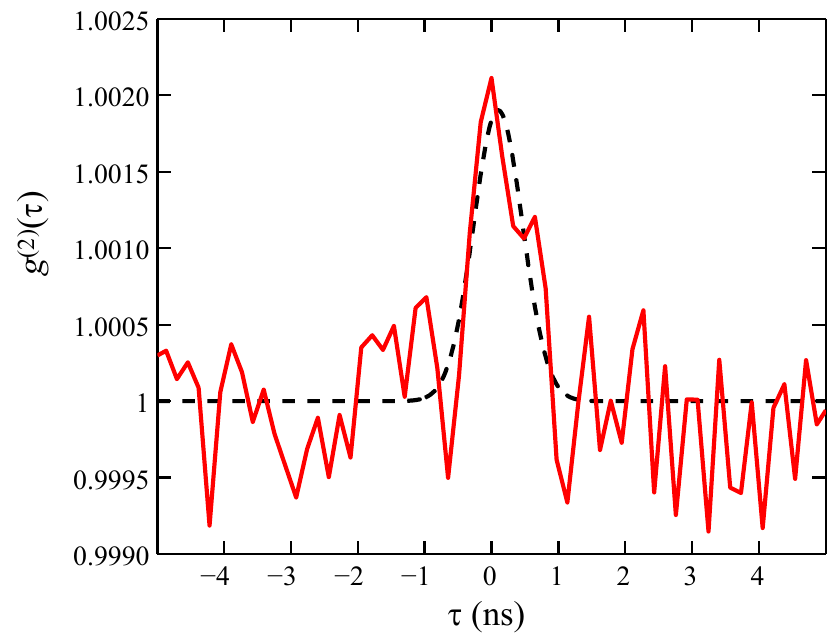}
        \caption{$\alpha$ CMi (Procyon).}\label{fig.g2b}
    \end{subfigure}
    \begin{subfigure}{0.66\columnwidth}
        \centering \includegraphics[width=\columnwidth]{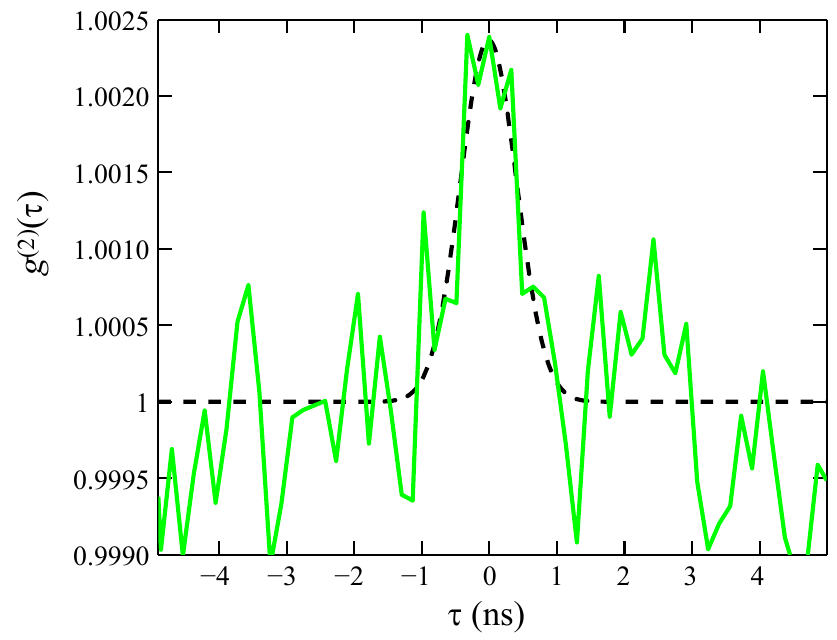}
        \caption{$\beta$ Gem (Pollux).}\label{fig.g2c}
    \end{subfigure}
    \caption{Temporal intensity correlation function $\gtau$ measured on three different stars. The Gaussian fit allows us to extract the contrast (reported in the Table~\ref{tab.Results}) and the FWHM. The fit window is $[-10 , 10]$~ns.}\label{fig.g2_Arcturus}
\end{figure*}

\begin{figure}
    \includegraphics[width=\columnwidth]{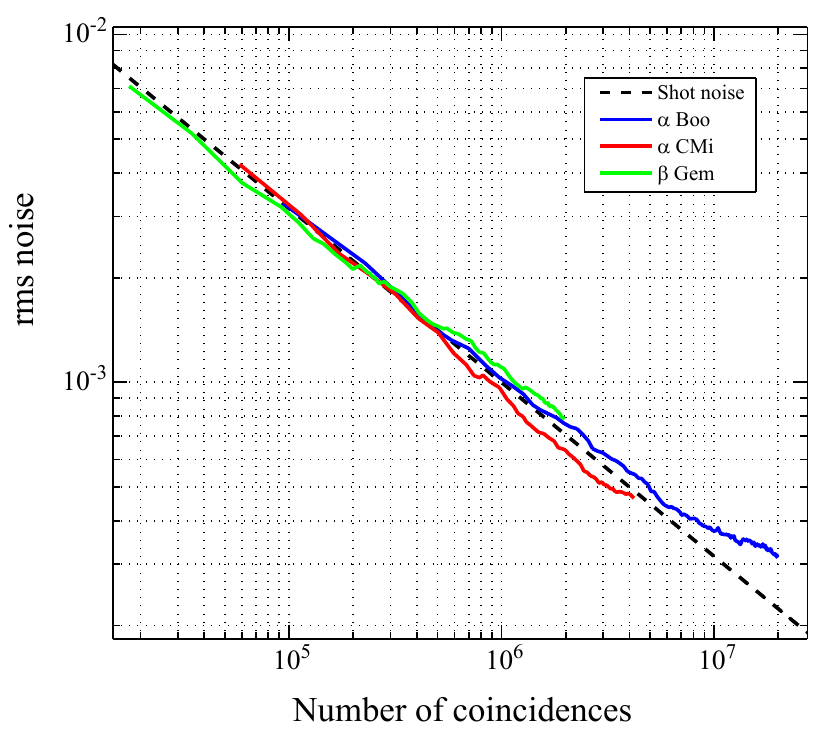}
    \caption{Noise on the correlation function as a function of the number of coincidences per time bin in the delay histogram. The dashed black line indicates the shot-noise expectation. For the three observed stars, the experimental noise is very close to the shot-noise level.}
    \label{fig.Noise}
\end{figure}

The second, and main result, is the observation of the intensity temporal correlation.
Fig.~\ref{fig.g2a} depicts the intensity correlation function measured on $\alpha$~Boo with a total integration time of almost $12$~hours over two consecutive nights. The bunching peak at $\tau = 0$ is well visible, with an amplitude of about $2\times 10^{-3}$ above one, as expected, and a residual rms noise of $3\times 10^{-4}$, slightly above the shot noise limit (see Fig.~\ref{fig.Noise}). A Gaussian fit to the data yields a measurement of the contrast $\mathcal{C} = (1.81 \pm 0.19) \times 10^{-3}$, and a FWHM of $700 \pm 90$~ps, consistent with the expectations.

We repeated the measurement with two other stars with slightly lower magnitudes. The results are plotted in Figs.~\ref{fig.g2b} and ~\ref{fig.g2c}. Apart from a different amount of noise, we obtained similar results, which are  summarized in Table~\ref{tab.Results}. The three measured contrasts are consistent between them and with the laboratory test.

\begin{table}
    \centering
    \caption{Summary of the observation results. $T$ is the total integration time, in hours and minutes. The flux $\mathcal{F}$ is the number of detected photon counts per second and per detector, averaged over the total integration time. The contrast $\mathcal{C} = g^{(2)}(0)-1$ is the value of the correlation at zero delay given by the amplitude of the Gaussian fit, its uncertainty is the $1\sigma$ confidence interval of the fit (see Fig.~\ref{fig.g2_Arcturus}). The noise is the rms noise on the data evaluated in the wings of the correlation function, where $\gtau \simeq 1$ (see Figs.~\ref{fig.g2_Arcturus} and~\ref{fig.Noise}).}
    \label{tab.Results}
    \begin{tabular}{lrccc}
        \hline
        Star & $T$ & $\mathcal{F}$ ($\times 10^6$ cps) & $\mathcal{C}$
        $(\times 10^{-3})$ & noise $(\times 10^{-3})$ \\
        \hline
        $\alpha$ Boo & 11:55' & 2.29 & $1.81 \pm 0.19$ & 0.32 \\
        $\alpha$ CMi &  4:35' & 1.44 & $1.90 \pm 0.28$ & 0.46 \\
        $\beta$ Gem  &  6:50' & 0.85 & $2.38 \pm 0.43$ & 0.78 \\
        \hline
    \end{tabular}
\end{table}

For the three stars, we show in Fig.~\ref{fig.Noise} the noise as a function of the number $N$ of coincidences per time bin recorded in the delay histogram. We essentially see a $N^{-1/2}$ scaling corresponding to the statistical uncertainty of a Poisson random variable, namely the `shot noise'. In this case, the SNR is given by~\citep{HBT:1967a}
\begin{equation}
\mathrm{SNR} = \alpha n_\mathrm{p} A \sqrt{\frac{\Delta f t}{2}},\label{eq.SNR2}
\end{equation}
where $n_\mathrm{p}$ is the number of photons per unit area, per unit time and per optical frequency, $A$ is the collecting area of the telescope, $\alpha$ the detection efficiency, which takes into account the APDs quantum efficiency as well as the injection efficiency and various losses, $\Delta f$ the electronic bandwidth, limited in our case by the APDs jitter, and $t$ the total integration time. Note that the SNR in Eq.\,(\ref{eq.SNR2}) is independent of the bandwidth of the optical filter.


One key element to observe photon bunching on stars is our capability to be shot noise limited, which allows to increase the SNR ratio above one by simply increasing the integration time. Shot noise limitation has been obtained by subtracting the TDC spurious correlation from our data (see Sec.\,\ref{sec.TDC}). Indeed, if this procedure is not applied, the noise is limited by the TDC to 1\%. With a contrast of about $10^{-3}$, the SNR would saturate to $10^{-1}$, whatever the total integration time is, making the photon bunching unobservable. This is actually the typical SNR obtained on different stars by Zampieri and coworkers in Asiago  ~\citep{Zampieri:2016}.

In summary, the measurements on the stars gave results perfectly consistent with the laboratory measurement, despite the fluctuations of the photon flux due to atmospheric turbulence. This demonstrates the robustness of this technique against turbulence, which can hamper amplitude interferometry.

\subsection{Discussion}\label{sec.g2_discussion}

It is also interesting to compare the performance of our measurement with the one of the Narrabri observatory. We can use $\alpha$~CMi for this purpose, as it has been measured in the two experiments. In ref.~\cite{HBT:1967b}, the contrast of the correlation measured on $\alpha$~CMi with the shortest baseline is reported with a relative $1\sigma$ uncertainty of 8.7\% ($SNR = 11.5$), with an observing time of 13.9~h. In our case, we have a relative uncertainty on the contrast of 14.7\% ($SNR = 6.8$) with an observing time of only 4.6~h and a total collection area approximately 85 times smaller (the Narrabri observatory used two 6.5~m collectors, whilst we use a 1~m telescope). The relatively good performance of our measurement is due to our larger electronic bandwidth, to the better quantum efficiency of the detectors, the fact that we do not have any loss of correlation due to the separation of the telescopes (the minimal baseline of the Narrabri interferometer was 9.5~m), and our working wavelength in the red ($\lambda_0 = 7800$~{\AA}), which is more adapted than the blue (4430~{\AA}) for F spectral type stars such as $\alpha$~CMi. However we lose photons because of the unperfect efficiency of injection in the fiber and because of the way we compute the correlation function~(see Sec.~\ref{sec.TDC}). These two limitations can be overcome by using a tip-tilt corrector (or even better, adaptive optics) to optimize and stabilize the light injection into the fiber and by using a better TDC or a more evolved data processing to exploit all detected photons.

\section{Conclusion}

In this paper we have reported the results of temporal intensity interferometry observations on three stars using a 1~m optical telescope. To our knowledge this is the first successful on sky experiments, aside from those on the sun, following the pioneering work of Hanbury Brown \emph{et al}. in the 1960s. Based on modern fiber transport of light, adapted to the aperture of the telescope at its Cassegrain focus, to feed photon-counting APDs, with special care taken to avoid dead-time detection and any correlation artefact, we attained a sensitivity comparable to that obtained by HBT, with a collection area almost two orders of magnitude lower and a shorter integration time.

Based on the simple assumption that SNR scales with the collecting area of the telescopes (see Eq.\,(\ref{eq.SNR2})), a 10~m class facility would enable us to perform measurements of similar accuracies on stars with 5 more magnitudes, using the same focal instrument in the same observing conditions as for experiment carried on the Calern 1~m telescope. Another factor of 3.5 in the SNR without losing any photons (and thus 1.5 in the observable magnitude) is within reach by using photodetectors with higher temporal resolution, thus increasing the electronic bandwidth. APDs with 40~ps timing resolution are indeed routinely used in quantum optics. Those detectors often have a very small surface, and a prerequisite is thus to inject the light in a single-mode fiber, which is doable with adaptive optics~\citep{Perrin:2006,LeBouquin:2011}. Along with other possible improvements (better observing site, better mirror reflectivity at other wavelengths, longer observing time, possibility to run parallel measurements on the two polarization modes of the incoming light and with arrays of detector on spectrally dispersed light as proposed by \cite{Trippe:2014}, etc.), magnitudes on the order of 10 seem within reach with the current technology. Thus, our successful preliminary observation paves the way to several developments.

First, using the single-telescope configuration described in this paper, we can measure the second order temporal coherence of light $\gtau$. By selecting one emission line with adequate narrow bandpass filters, one can deduce its spectral width from the coherence time measured in the $\gtau$ function~\citep{Phillips:1967}. The coherence time being inversely proportional to the spectral width, this technique is particularly suited to study very narrow emission lines that are not resolved by standard spectroscopic techniques~\citep{Dravins:2008a}. Beyond the measurement of spectral width, we can also characterize non-thermal emission lines, such as the so-called astrophysical lasers~\citep{Johansson:2007,Letokhov:2009,Dravins:2008c}. In particular, a drop of the contrast of the $\gtau$ function would witness the signature of gain saturation along the propagation and the subsequent increase of coherence~\citep{Chung:1980}, and a flat correlation function would reveal completely coherent light, such as a true single-mode laser~\citep{Arecchi:1966,Loudon:book}.
On a longer term, the foreseeable 30-40~m optical telescopes such as the TMT and the E-ELT, with collecting areas 1000 to 1500 times the power of a 1~m telescope, will represent the true playground to detect such phenomena.

Second, we can extend our setup to the HBT configuration and carry on spatial intensity interferometry using two telescopes. This can be achieved in a first step using the two 1~m OCA-Calern telescopes on an E-W baseline of 15~m, that can be readily extended to 100~m using the 1.5~m laser-ranging telescope available at OCA-Calern. Finally, an extension to three-telescope intensity interferometry could be conducted. In addition, the performances of these interferometers could greatly benefit from spectral multiplexing techniques using APD arrays~\citep{Trippe:2014}. If successful this will open a new perspective for optical long baseline interferometry at visible short wavelengths down to 0.4~$\mu$m and within atmospheric transmission limits depending on the observing site.

Within its intrinsic sensitivity limitations, the simplicity of our intensity interferometry instrument opens some unique opportunities with existing optical telescopes nowadays grouped on a same observatory, as proposed by several groups~\citep{Zampieri:2016,Pilyavsky:2017}, or using Cherenkov telescope arrays~\citep{Dravins:2016} for imaging interferometry in the visible at sub-mas resolutions.

\section*{Acknowledgements}

We thank Aur\'elien Eloy and Patrizia Weiss for their participation to the observations, Robert Sellem for insightful discussions on TDCs, Armando Domiciano de Souza for careful reading of the manuscript, S\'ebastien Tanzilli, Olivier Alibart, Michel Lintz, Philippe Bendjoya and Lyu Abe for fruitful discussions, and Gerard van Belle for useful comments on the manuscript. AD acknowledges the financial support of Campus France (Program No. CF-PRESTIGE-18-2015). We also acknowledge funding from the D\"oblin Federation and P.~Berio, head of Lagrange technical staff.



\label{lastpage}
\end{document}